\documentclass[conference]{IEEEtran}
\IEEEoverridecommandlockouts

\usepackage{color}
\usepackage{cite}
\usepackage{url}

\ifCLASSINFOpdf
\else
\fi
\usepackage{caption}
\usepackage{amsfonts,amssymb}
\usepackage{graphicx}
\usepackage{epstopdf}
\usepackage{mathrsfs}
\usepackage{amsmath}
\usepackage{latexsym}
\usepackage{booktabs}
\usepackage{subfigure}
\usepackage{algorithm, algorithmic}
\usepackage{bbm}
\usepackage[amsmath,thmmarks]{ntheorem}
\usepackage{bm}
\begin{document}
%
\title{\huge Gated Recurrent Units Learning for Optimal Deployment of Visible Light Communications Enabled UAVs}




%
\author{\IEEEauthorblockN{Yining Wang\IEEEauthorrefmark{1},
Mingzhe Chen\IEEEauthorrefmark{2}$^,$\IEEEauthorrefmark{3},
Zhaohui Yang\IEEEauthorrefmark{4},
Xue Hao\IEEEauthorrefmark{1},
Tao Luo\IEEEauthorrefmark{1},
and Walid Saad\IEEEauthorrefmark{5}}
\IEEEauthorblockA{\small \IEEEauthorrefmark{1}Beijing Laboratory of Advanced Information Network, Beijing University of Posts and Telecommunications,\\
 Beijing, China 100876, Emails: \protect\url{wyy0206@bupt.edu.cn}, \protect\url{haoxueisme@bupt.edu.cn}, \protect\url{tluo@bupt.edu.cn}.\\
\IEEEauthorrefmark{2}Department of Electrical Engineering, Princeton University, \\
and also with the Chinese University of Hong Kong, Shenzhen, China, Email: \protect\url{mingzhec@princeton.edu}.\\
\IEEEauthorrefmark{3}The Future Network of Intelligence Institute, The Chinese University of Hong Kong, Shenzhen, China.\\
\IEEEauthorrefmark{4}Centre for Telecommunications Research, Department of Engineering, Kings College London, \\
WC2B 4BG, UK, Email: \protect\url{yang.zhaohui@kcl.ac.uk}.\\
\IEEEauthorrefmark{5}Wireless@VT, Bradley Department of Electrical and Computer Engineering,\\
Virginia Tech, Blacksburg, VA, USA, Email: \protect\url{walids@vt.edu}.
\thanks{This work was supported in part by the National Natural Science Foundation of China under Grant 61571065, by the U.S. National Science Foundation under Grants CNS-1909372 and IIS-1633363, and by grants No. ZDSYS201707251409055, No. 2017ZT07X152, No. 2018B030338001, and No. 2018YFB1800800.}
}\vspace*{-3em}
}


\maketitle

\begin{abstract}
In this paper, the problem of optimizing the deployment of unmanned aerial vehicles (UAVs) equipped with visible light communication (VLC) capabilities is studied. In the studied model, the UAVs can simultaneously provide communications and illumination to service ground users. Ambient illumination increases the interference over VLC links while reducing the illumination threshold of the UAVs. Therefore, it is necessary to consider the illumination distribution of the target area for UAV deployment optimization. This problem is formulated as an optimization problem whose goal is to minimize the total transmit power while meeting the illumination and communication requirements of users. To solve this problem, an algorithm based on the machine learning framework of gated recurrent units (GRUs) is proposed. Using GRUs, the UAVs can model the long-term historical illumination distribution and predict the future illumination distribution. In order to reduce the complexity of the prediction algorithm while accurately predicting the illumination distribution, a Gaussian mixture model (GMM) is used to fit the illumination distribution of the target area at each time slot. Based on the predicted illumination distribution, the optimization problem is proved to be a convex optimization problem that can be solved by using duality. Simulations using real data from the Earth observations group (EOG) at NOAA/NCEI show that the proposed approach can achieve up to 22.1\% reduction in transmit power compared to a conventional optimal UAV deployment that does not consider the illumination distribution. The results also show that UAVs must hover at areas having strong illumination, thus providing useful guidelines on the deployment of VLC-enabled UAVs.
\end{abstract}


%
\IEEEpeerreviewmaketitle

\section{Introduction}
Deploying unmanned aerial vehicles (UAVs) for wireless networking is a flexible and cost-effective approach to providing on-demand communications \cite{MingzheChen}. However, the limited bandwidth of radio frequency bands and limited energy of UAVs confine the applicability of UAV-enabled wireless networks.
This challenge can be addressed by equipping UAVs with visible light communication (VLC) capabilities. Indeed, VLC has recently attracted attention due to its large license-free bandwidth and high energy efficiency. For instance, a VLC system that uses light-emitting diodes (LEDs) to transmit wireless signals can provide both illumination and communication services. Moreover, the altitude of the UAVs ensures the line of sight channel for VLC. Therefore, using VLC can be a promising approach to provide energy-efficient UAV communications with sufficiently available bandwidth. However, deploying VLC-enabled UAVs also faces many challenges that range from illumination interference detection and prediction to UAV deployment optimization and energy efficiency.

The existing literature such as in \cite{ICC, Mohammad, JSAC, Liquid,twouser} has studied a number of problems related to UAV deployment.
In \cite{ICC}, the authors proposed an optimization framework in UAV-assisted backscatter networks to identify the trade-off between UAV's altitude, number of backscatter devices and backscatter coefficients.
The authors in \cite{Mohammad} derived the average coverage probability and the system sum-rate as a function of the UAV altitude and the number of users.
However, the works in \cite{ICC} and \cite{Mohammad} only consider the altitude of UAVs without optimizing their locations.
In \cite{JSAC}, the authors studied the optimal UAVs' locations based on the prediction of human behavior so as to optimize the quality-of-experience of wireless devices.
The authors in \cite{Liquid} studied the use of neural network based learning algorithm to optimize the performance of the UAV based wireless networks.
The work in \cite{twouser} optimized the locations of UAVs in two-user broadcast channel and proved that the UAVs must hover around the user who has a large rate requirement.
However, the works in \cite{JSAC, Liquid, twouser} ignored the energy efficiency of UAVs in optimizing the deployment of UAVs.
Moreover, all of the existing works such as in \cite{ICC, Mohammad, JSAC, Liquid, twouser} are over limited capacity radio frequency bands which may not allow the UAVs to meet the high data rate demands of ground users.
Instead, VLC-enabled UAVs can be considered to provide high speed communications \cite{yang}.
However, nighttime illumination such as vehicle lights, street lights and building lights causes strong interference to VLC link.
Therefore, it is necessary to analyze the illumination distribution of the service areas so as to optimize the deployment of VLC-enabled UAV.

The main contribution of this work is a novel framework for dynamically optimizing the locations of VLC-enabled UAVs based on accurate predictions of the illumination distribution of a given area. To the best of our knowledge, this is the first work that \emph{studies the use of the predictions of the illumination distribution to provide a power-efficient deployment of VLC-enabled UAVs}. Our key contributions include:
\begin{itemize}
\item We consider a VLC-enabled UAV network, in which the UAVs must find their optimal locations by predicting the distribution of ambient lighting so as to provide illumination as well as communication services to ground users. This problem is formulated as an optimization problem whose goal is to minimize the total transmit power of UAVs under both illumination and communication constraints.
\item To solve this optimization problem, we first use a Gaussian mixture model (GMM) to fit the illumination distribution, which can be obtained using an expectation-maximization (EM) algorithm. Based on derived illumination distribution, a gated recurrent units (GRUs) based prediction algorithm is proposed to predict future illumination distribution. The proposed prediction algorithm can model the temporal characteristics of the long-term historical illumination distribution thus enabling the UAVs to predict future illumination distributions.
\item Given the predicted illumination distribution, the original optimization problem is proved to be a convex problem which is then solved using duality.
\item Simulation results show that the proposed approach can achieve up to 22.1\% reduction in terms of transmit power compared to a conventional optimal UAV deployment without considering illumination distribution. Furthermore, simulations indicate that UAVs should hover over areas with strong illumination.
\end{itemize}

The rest of this paper is organized as follows. The system model and the problem formulation are described in Section \ref{sec:2}. The use of the GMM to model illumination distribution, GRU-based algorithm for illumination prediction, and the optimization of the UAV deployment are proposed in Section \ref{sec:3}. In Section \ref{sec:4}, the numerical results are presented and discussed. Finally, conclusions are drawn in Section \ref{sec:5}.

\section{System Model and Problem Formulation}
\label{sec:2}
\begin{figure}
\centering
\setlength{\belowcaptionskip}{-0.45cm}
\includegraphics[width=6.5cm]{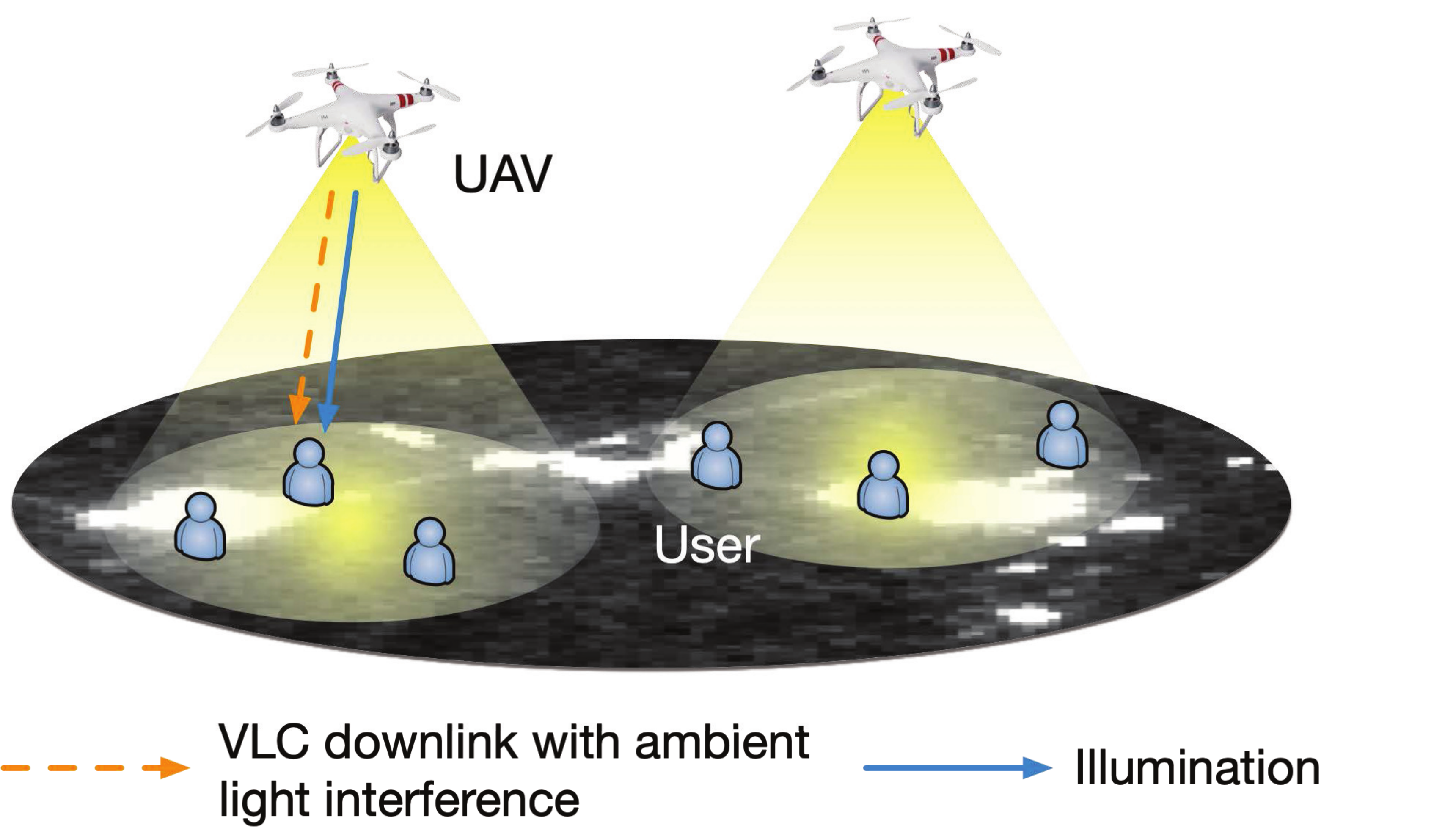}
\caption{The architecture of a cellular network that consists of UAVs and users.}
\label{fig1}
\vspace{-0.1cm}
\end{figure}

Consider a wireless network composed of a set $\cal{D}$ of $\emph{D}$ VLC-enabled UAVs that serve a set $\cal{U}$ of $\emph{U}$ ground users distributed over a geographical area $\mathcal{A}$. The UAVs provide downlink transmission and illumination simultaneously as shown in Fig. \ref{fig1}. Hereinafter, we use $\emph{aerial cell}$ to refer to the service area of each UAV. We assume that each UAV $i \in \mathcal{D}$ only serves the users located in its aerial cell ${\cal A}_i$.

Given a UAV $ i \in {\cal D}$ located at $({x_i},{y_i},H)$ and a ground user $j \in {\cal U}$ located at $({x_j},{y_j}) \in {\cal A}$, the channel gain of the VLC link between UAV $i$ and user $j$ can be given by \cite{KomineFundamental}:
\vspace{-0.1cm}
\begin{equation}
\vspace{-0.1cm}
{h_j}(x_i,y_i) = \left\{ {\begin{array}{*{20}{l}}
{\!\frac{{(m + 1)S}}{{2\pi d_{ij}^2}}g(\psi ){{\cos }^m}(\phi )\cos \left( \psi  \right),0\leqslant \psi \leqslant \Psi_c,}\\
{\!\;\;\;\;\;\;\;\;\;\;\;\;\;\;\;\;\;\;\;0\;\;\;\;\;\;\;\;\;\;\;\;\;\;\;\;\;\;\;\;\;\;,\psi > {\Psi _c},}
\end{array}} \right.
\label{channel_model}
\end{equation}
where $S$ is the detector area and $d_{ij} = \sqrt {{{({x_j} - {x_i})}^2} + {{({y_j} - {y_i})}^2} + H^2} $ is the distance between UAV $i$ and ground user $j$. $m =  - \ln 2/\ln (\cos {\Phi _{1/2}})$ is the Lambert order with ${\Phi _{1/2}}$ being the transmitter semiangle (at half power). $\psi$ and $\phi$ represent the angle of incidence and irradiance, respectively. As such, $\cos \phi  = \cos \psi  = \frac{H}{d_{ij}}$. ${{\Psi _c}}$ is the receiver field of vision (FOV) semiangle. The gain of optical concentrator $g(\psi )$ is defined as:
\vspace{-0.1cm}
\begin{equation}
\vspace{-0.1cm}
g(\psi ) = \left\{ {\begin{array}{*{20}{l}}
{\frac{{{n_e}^2}}{{{{\sin }^2}{\Psi _c}}},\;0 \leqslant {\psi} \leqslant {\Psi _c},}\\
{\;\;\;\;0,\;\;\;\;\;{\psi} > {\Psi _c}},
\end{array}} \right.
\end{equation}
where $n_e$ represents the refractive index.

For static user $j$ located at $\left(x_j,y_j\right) \in {\cal A}_i$, the channel capacity at time $t$ can be given by:
\vspace{-0.1cm}
\begin{equation}
\vspace{-0.1cm}
{C_{ij,t}} = \frac{1}{2}{\log _2}\left( {1 + \frac{e}{{2\pi }}{{\left( {\frac{{\xi {P_{ij,t}}{h_j}(x_i,y_i)}}{{{n_w} + {I_t}(x_j,y_j)}}} \right)}^2}} \right),\label{eq:capacity}
\end{equation}
where $\xi$ is the illumination target, $P_{ij,t}$ is the transmit power of UAV $i$ for user $j$ at time $t$, and ${n_w}$ represents the standard deviation of the additive white Gaussian noise.
We define the illumination distribution of the target area as $I_t(x,y)$ that will be specified in Section III. Note that the illumination distribution $I_t(x,y)$ is the luminance of the given area and it is also the interference over the VLC transmission links caused by illumination due to human activity. Here, ${I_t}(x_j,y_j)$ is the interference over the VLC link between the UAV and the user located at $(x_j,y_j)$.

Due to the limited energy of UAVs, their deployment must be optimized to minimize the transmit power while satisfying the data rate and illumination requirements of users. As done in \cite{MBS}, we do not consider the mobility energy consumption of the UAVs.

\subsection{Problem Formulation}
To formalize the deployment problem, we must first determine the minimum transmit power that each UAV $i$ uses to meet the data rate and illumination requirements of its associated users. To satisfy the data rate constraint $R_j$ of each user $j$ located at $(x_j,y_j) \in {\cal A}_i$, the required power of UAV $i$ at time $t$ is:
\vspace{-0.1cm}
\begin{equation} \label{power}
\vspace{-0.1cm}
P_{ij,t} = \frac{{({n _w} + {I_t}(x_j,y_j))\sqrt {\frac{{2\pi }}{e}({2^{2{R_j}}} - 1)} }}{{\xi {h_j}(x_i,y_i)}}.
\end{equation}
A UAV can successfully satisfy all the users’ requirements once the user that has the maximum power requirement is satisfied. Therefore, the minimum transmit power of UAV $i$ satisfying the data rate requirements of its associated users is given by:
\vspace{-0.1cm}
\begin{equation}\label{power_min}
\vspace{-0.1cm}
P_{i,t}^\textrm{min} = \max \{ {P_{ij,t}}\} ,\forall j \in {{\cal U}_i},
\end{equation}
where ${\cal U}_i$ is the set of users associated with UAV $i$.

Given this system model, our goal is to find an effective deployment of UAVs that meets the data rate requirements of each user while minimizing the transmit power of the UAVs. This problem involves predicting the illumination and adjusting the locations as well as the transmit powers of the UAVs. The optimization problem is formulated as follows:
\vspace{-0.1cm}
\begin{subequations}\label{optimal_problem}
\begin{align}\tag{6}
&{\mathop {\min }\limits_{{{I_t}(x,y)},{x_i},{y_i}} \!\sum\limits_{i \in {\cal{D}}} {{P_{i,t}}} ,}\\
&{\;\;\;\;\;\;{\rm{s}}.{\rm{t}}.\;\;\;\;\;{\xi {P_{i,t}}{h_j}(x_i,y_i)}\! + \!{I_t(x_j,y_j)}\! \geqslant \! \eta _r,\! \forall i \! \in \! {\cal{D}},\! \forall j \! \in \! {\cal{U}}_i},\\
&{\;\;\;\;\;\;\;\;\;\;\;\;\;\;\;\;{{P}_{i,t}}\! \geqslant \! P_{i,t}^{\min },\! \forall i \! \in \! {\cal{D}},}
\vspace{-0.1cm}
\end{align}
\end{subequations}
where ${\eta _r}$ denotes the illumination demand and ${\xi {P_{i,t}}{h_i}(x,y)}$ is the illumination of UAV $i$ at time $t$ \cite{illumination}. (6a) indicates that the UAV needs to provide illumination to meet the illumination threshold of each user $j$. (6b) indicates that the transmit power of UAV $i$ should satisfy the data rate requirements of its associated users from (\ref{power_min}). Here, we note that ambient illumination increases the VLC interference while reducing the illuminance threshold. Since the illumination distribution is time-varying, it is necessary to predict the illumination distribution of the target area to deploy the UAVs at the beginning of each time interval. Hence, we introduce a machine learning algorithm to predict the illumination distribution of the service area. Based on the prediction, UAVs are deployed according to the solution of (\ref{optimal_problem}) which remain unchanged during each prediction period.
\section{Machine Learning for Illumination Prediction and UAV Deployment}
\label{sec:3}
Here, we first introduce the use of GMM to model the illumination distribution. Then, we propose a GRU-based machine learning algorithm for predicting the illumination distribution. The proposed learning algorithm enables the UAVs to analyze the relationship among historical illumination distributions, to predict the future illumination distribution. Based on the prediction, the UAVs can be optimally deployed.

\subsection{GMM Fitting of Illumination Distribution}
Since illumination is caused by human activities such as business and industrial operation, the illumination distribution of a given area is centered on several small areas. Therefore, we assume that the illumination distribution of area ${\cal A}$ at time $t$ follows a GMM given by:
\begin{equation}\label{illumination_distribution}
\begin{split}
&{I_t}(x,y) = A_t \times \\
&\left(\sum\limits_{k = 1}^K {{w_{k,t}}\exp \left( - \frac{{{{(x - {\mu_{x_{k,t}}})}^2}}}{{2\sigma _{{x_{k,t}}}^2}} - \frac{{{{(y - {\mu_{y_{k,t}}})}^2}}}{{2\sigma _{{y_{k,t}}}^2}}\right)} \right),
\end{split}
\end{equation}
where ${A_t}$ is the GMM amplitude and $K$ is the number of Gaussian components with $\sum\limits_{k = 1}^K {{w_k} = 1}$ with ${w_k} \in [0,1]$ being the weight of each component. For each Gaussian component $k$, $({\mu_{x_{k,t}}},{\mu_{x_{k,t}}})$ is the central coordinate while $\sigma _{{x_{k,t}}}$ and $\sigma _{{y_{k,t}}}$ represent the standard deviation on the $x$ and $y$ axes, respectively. The parameters related to Gaussian component $k$ at time $t$ can be represented as a vector $\bm{q}_{k,t}=\left[{w_{k,t}},{\mu_{x_{k,t}}},{\mu_{y_{k,t}}},{\sigma _{{x_{k,t}}}},{\sigma _{{y_{k,t}}}}\right]$. Given (\ref{illumination_distribution}), the spatial features of illumination distribution at time $t$ can be represented by a vector, which can be given by:
\vspace{-0.1cm}
\begin{equation}
\begin{aligned}
\vspace{-0.2cm}
\bm{q}_t &{= \left[A_t; \bm{q}_{1,t}; \bm{q}_{2,t}; \ldots; \bm{q}_{K,t}\right]^ \mathsf{T}}\\
&{=\! \left[A_t, {w_{1,t}},{\mu_{x_{1,t}}},{\mu_{y_{1,t}}},{\sigma _{{x_{1,t}}}},{\sigma _{{y_{1,t}}}}, \right.}\\
&{\phantom{=\;\;}\left. {w_{2,t}},{\mu_{x_{2,t}}},{\mu_{y_{2,t}}},{\sigma _{{x_{2,t}}}},{\sigma _{{y_{2,t}}}}, \right.}\\
&{\phantom{=\;\;}\left.\ldots,{w_{K,t}},{\mu_{x_{K,t}}},{\mu_{y_{K,t}}},{\sigma _{{x_{K,t}}}},{\sigma _{{y_{K,t}}}} \right]^ \mathsf{T}.}
\end{aligned}
\end{equation}
The optimal vector $\bm{q}_t$ that is used to model the illumination distribution can be determined using the EM algorithm \cite{EM}.

\vspace{-0.1cm}
\subsection{Illumination Distribution Prediction}
Next, we introduce the use of GRUs \cite{GRU} for the prediction of the illumination distribution. GRUs are extensions of conventional recurrent neural networks (RNNs)\cite{tutorial}. GRUs can effectively solve the gradient vanishing and the gradient exploding problem in long-term memory RNNs. Due to interconnected neurons at hidden layers and their internal gating mechanisms, GRUs can model the temporal characteristics of the long-term illumination distribution. In addition, GRUs can dynamically update the model based on the current illumination distribution due to the variable-length recurrent structure, hence, GRUs enable the UAVs to predict future illumination distribution.

A GRU-based prediction algorithm consists of three components: a) input, b) output, and c) GRU model. The key components of our GRU-based prediction approach are:
\begin{itemize}
\item \emph{Input}: The input of the GRU-based prediction algorithm is a matrix $\boldsymbol{Q}$ that represents the time series data of illumination distribution, which can be given by $\boldsymbol{Q} = \left[\bm{q}_1, \cdots ,\bm{q}_T\right],$ where $T$ is the length of the time series.
\item \emph{Output}: The output of the GRU-based prediction algorithm is a vector $\bm{q}_{T+1}$, that represents the illumination distribution at time slot $T+1$.
\item \emph{GRU model}:  A GRU model is used to approximate the function between the input $\boldsymbol{Q}$ and output $\bm{q}_{T+1}$, thus building a relationship between historical illumination distribution and future illumination distribution. A GRU model is essentially a dynamic neural network that consists of an input layer, a hidden layer, and an output layer. The hidden states $\bm{h}_t$ of the units of the in hidden layer at time $t$ are used to store information related to the illumination distribution from time slot $1$ to $t$. For each time $t$, the hidden states $\bm{h}_t$ of the GRU are updated based on the input $\bm{q}_t$ and $\bm{h}_{t-1}$. Next, we introduce how to update the hidden state $h_t^j$ of hidden unit $j$ given a new illumination distribution $\bm{q}_t$.

\setlength\parindent{1em}At each time slot $t$, the hidden state $h_t^j$ is determined by two gates: reset gate $r_t^j$ and update gate $z_t^j$. First, the reset gate $r_t^j$ is used to determine the historical illumination distribution information retained in the candidate hidden state ${\tilde{h}_t^j}$, which can be given by:
\vspace{-0.2cm}
\begin{equation}
\vspace{-0.1cm}
r_t^j = \sigma ({[{{\boldsymbol{W}}_r}{\bm{q}_t}]_j} + {[{{\boldsymbol{U}}_r}{\bm{h}_{t - 1}}]_j}),
\end{equation}
where $\sigma (\cdot)=\frac{1}{{1 + {e^{ - (\cdot)}}}}$ is the logistic sigmoid function and ${[ \cdot ]_j}$ is element $j$ of a vector. ${{\boldsymbol{W}}_r} \in {\mathbb{R}^{{D_q} \times {D_h}}}$ and ${{\boldsymbol{U}}_r} \in {\mathbb{R}^{{D_h} \times {D_h}}}$ represent the weight matrices of reset gate, where $D_q= 5K+1$ is the length of the input ${\bm{q}_t}$ and $D_h$ is the number of the units in hidden layer. Based on the value of the reset gate $r_t^j$, the candidate hidden state ${\tilde{h}_t^j}$ that is used to combine the input illumination distribution $\bm{q}_t$ with the previous memory ${\bm{h}_{t - 1}}$ is given by:
\vspace{-0.1cm}
\begin{equation}\label{reset}
\vspace{-0.1cm}
{\tilde{h}_t^j} = \tanh \left({[{{\boldsymbol{W}}_{\tilde{h}}}{\bm{q}_t}]_j} + {[{{\boldsymbol{U}}_{\tilde{h}}}({\bm{r}_t} \odot {\bm{h}_{t - 1}})]_j}\right),
\end{equation}
where ${\bm{r}_t} \in  {\mathbb{R}^{D_h}}$ is a reset gate vector at time $t$ and $\odot$ is an element-wise multiplication. For example, given two vectors ${\bm{x}}=(a,b)$ and ${\bm{y}}=(c,d)$, ${\bm{x}} \odot {\bm{y}}=(ac,bd)$. ${{\boldsymbol{W}}_{\tilde{h}}} \in {\mathbb{R}^{{D_q} \times {D_h}}}$ and ${{\boldsymbol{U}}_{\tilde{h}}} \in {\mathbb{R}^{{D_h} \times {D_h}}}$ represent the hidden state weight matrices.

Similarly, the update gate ${z_t^j}$ is used to decide the size of the information stored in the candidate hidden state to update the hidden state $h_t^j$, which can be given by:
\vspace{-0.1cm}
\begin{equation}
{z_t^j} = \sigma \left({[{{\boldsymbol{W}}_z}{\bm{q}_t}]_j} + {[{{\boldsymbol{U}}_z}{\bm{h}_{t - 1}}]_j}\right),
\end{equation}
where ${{\boldsymbol{W}}_z} \in {\mathbb{R}^{{D_q} \times {D_h}}}$ and ${{\boldsymbol{U}}_z} \in {\mathbb{R}^{{D_h} \times {D_h}}}$ represent the weight matrices of the update gate. The actual hidden state $h_t^j$ of hidden unit $j$ is updated by:
\begin{equation}
h_t^j =  {z_t^j}{h_{t-1}^j} + (1 - {z_t^j}){\tilde{h}_t^j}.
\end{equation}

The proposed GRU model iteratively updates the hidden states to store the input $\boldsymbol{Q}$ until the hidden state of the current time $T$ is computed. The output layer of the GRU model will predict the illumination distribution at time ${T+1}$ based on the hidden state ${\bm{h}_T}$:
\vspace{-0.1cm}
\begin{equation}\label{output}
\vspace{-0.1cm}
\tilde{\bm{q}}_{T+1} =  {{\boldsymbol{W}}_o}{\bm{h}_T},
\end{equation}
where ${\boldsymbol{W}}_o \in {\mathbb{R}^{{D_h} \times {D_q}}}$ is the output weight matrix. In fact, (\ref{output}) is used to build the relationship between output $\bm{q}_{T+1}$ and the hidden state ${\bm{h}_T}$ that stores the information of input $\boldsymbol{Q}$.

To build this relationship, a batch gradient descent approach is used to train the weight matrices which are initially generated randomly via a uniform distribution. The update rule of the gradient descent approach is:
\vspace{-0.1cm}
\begin{equation}\label{weight}
\vspace{-0.1cm}
\begin{split}
{\boldsymbol{W}}_n^{i+1} &= {\boldsymbol{W}}_n^i - \alpha \nabla E({\boldsymbol{W}}_n),\\
{\boldsymbol{U}}_m^{i+1} &= {\boldsymbol{U}}_m^i - \alpha \nabla E({\boldsymbol{U}}_m),
\end{split}
\end{equation}
where $\alpha$ is the learning rate, $n \in \left \{ r, z, \tilde{h}, o \right \}$ and $m \in \left\{r, z, \tilde{h} \right\}$. $\nabla E({\bf{W}}_n) = \frac{{\partial E}}{{\partial {{\boldsymbol{W}}_n}}}$ and $\nabla E({\boldsymbol{U}}_m) = \frac{{\partial E}}{{\partial {{\boldsymbol{U}}_m}}}$ are the gradients of the loss function $E$ which is defined as:
\vspace{-0.1cm}
\begin{equation}\label{loss}
\vspace{-0.1cm}
{E} = \sum\limits_{t = 1}^{T-1} {\frac{1}{2}{({\bm{q}_{t+1}} - {\tilde{\bm{q}}_{t+1}})^2}} ,
\end{equation}
where ${\tilde{\bm{q}}_{t+1}}$ is the illumination distribution prediction and ${\bm{q}_{t+1}}$ is the actual illumination distribution at time $t+1$. The specific process of using the GRU-based prediction algorithm to predict the illumination distribution for each UAV $i$ is summarized in \textbf{Algorithm 1}.
\end{itemize}
\vspace{-0.2cm}

\begin{algorithm}[t]
\vspace{-0.05cm}
\footnotesize
\caption{GRU-based Prediction Algorithm for Illumination Distribution Prediction.}
\begin{algorithmic}[1]
\STATE \textbf{Input:} The time series illumination distribution $\boldsymbol{Q}$ of service area.
\STATE \textbf{Initialize:} ${\boldsymbol{W}_r}, {\boldsymbol{U}_r}, {\boldsymbol{W}_z}, {\boldsymbol{U}_z}, {{\boldsymbol{W}}_{\tilde{h}}}, {{\boldsymbol{U}}_{\tilde{h}}}, {\boldsymbol{W}_o}$ are initially generated randomly via a uniform distribution. The number of iterations $e$.
\FOR {$i = 1 \to e$}
\FOR {each time $t$}
\STATE Input ${\bm{q}_t}$ and ${\bm{h}_{t - 1}}$. Estimate the illumination distribution $\tilde{\bm{q}}_{T+1}$ at time $t+1$ based on (\ref{output}).
\ENDFOR
\STATE Calculate the loss $E$ based on (\ref{loss}).
\STATE Update the weight matrices based on (\ref{weight}).
\ENDFOR
\STATE \textbf{Output:} Prediction ${\bm{q}_{T+1}}$.
\end{algorithmic}
\label{algorithm_1}
\vspace{-0.05cm}
\end{algorithm}

\subsection{Optimization of the UAV Deployment}
Once the illumination distribution is predicted, the UAVs can determine their optimal deployment at the beginning of each time interval by solving the optimization problem defined in (\ref{optimal_problem}). As analyzed in Section \ref{sec:2}, a UAV only needs to consider the user with the maximum power requirement since, by doing so, the requirements of all other users will be automatically satisfied. Therefore, substituting (\ref{channel_model}) and (\ref{power_min}) into (\ref{optimal_problem}), we have:
\vspace{-0.5cm}
\begin{subequations}\label{optimal_problem2}
\begin{align}\tag{\theequation}
&{\mathop {\min \;}\limits_{{x_i},{y_i},P_{i,T + 1}} \;\sum\limits_{i \in {\cal{D}}} {{P_{i,T + 1}}} ,}\\
&{\;\;\;\;\;{\rm{s}}.{\rm{t}}.\;\;\;\;\;\;\;{P_{i,T + 1}} \geqslant lM{d_{ij}^{m+3}},\; \forall i \in {\cal{D}},\forall j \in {{\cal{U}}_i},}\\
&{\;\;\;\;\;\;\;\;\;\;\;\;\;\;\;\;\;{P_{i,T + 1}} \geqslant lN{d_{ij}^{m+3}}, \;\forall i \in {\cal{D}},\forall j \in {{\cal{U}}_i},}
\end{align}
\end{subequations}
where $l=\frac{2\pi}{{\xi (m + 1) S g(\psi ) H^{m+1} }}$, $M={{\eta _r} - {I_{T + 1}(x_j,y_j)}}$ and $N={({n_w} + {I_{T + 1}(x_j,y_j)})\sqrt {\frac{{2\pi }}{e}({2^{2{R_j}}} - 1)}}$.

Since the service area of each UAV does not overlap with the service areas of other UAVs, we ignore the interference caused by other UAVs. In consequence, problem \eqref{optimal_problem2} can be decoupled into multiple subproblems. For each UAV $i$, the location optimization subproblem can be formulated as:
\vspace{-0.2cm}
\begin{subequations}\label{optimal_problem3}
\begin{align}\tag{\theequation}
&{\mathop {\min \;}\limits_{{x_i},{y_i},P_{i,T + 1}} \;  {{P_{i,T + 1}}} ,}\\
&{\;\;\;\;\;\;{\rm{s}}.{\rm{t}}.\;\;\;\;\;\;{P_{i,T + 1}}^{\frac{2}{m+3}} \geqslant {a_j}{d_{ij}^2} , \forall j \in {{\cal{U}}_i},}
\end{align}
\end{subequations}
where $a_j= (\max\left\{lM,lN\right\})^{\frac{2}{m+3}}$.

Due to its convex objective functions and constraints, problem \eqref{optimal_problem3} is a convex problem, which can be optimally solved by using the dual method \cite{zhaohui}.
The Lagrange function of problem \eqref{optimal_problem3} will be:
\vspace{-0.2cm}
\begin{align}
\vspace{-0.3cm}
&\mathcal L={P_{i,T + 1}} +
\nonumber\\&
\sum_{j\in\mathcal U_i}  \lambda_j\left((({x_i} - {x_j}) ^2  +  ({y_i} - {y_j}) ^2  + H^2) a_j   -{P_{i,T + 1}}^{\frac{2}{m+3}}   \right),
\end{align}
where $\lambda_j$ is the dual variable associated with constraint $j$ in (\ref{optimal_problem3}a).

The optimal first-order conditions of (\ref{optimal_problem3}) will be:
\vspace{-0.15cm}
\begin{equation}\label{optimal_problem3eq0}
\vspace{-0.15cm}
\begin{split}
&\frac{\partial \mathcal L}{\partial P_{i,T + 1}}=1-\frac{2}{m+3}\sum_{j\in\mathcal U_i}  \lambda_j{P_{i,T + 1}}^{\frac{-m-1}{m+3}}  =0,
\end{split}
\end{equation}
\begin{equation}\label{optimal_problem3eq1}
\vspace{-0.2cm}
\begin{split}
&\frac{\partial \mathcal L}{\partial x_i}= 2\sum_{j\in\mathcal U_i}  \lambda_j a_j  ({x_i} - {x_j})=0,
\end{split}
\end{equation}
\begin{equation}\label{optimal_problem3eq2}
\vspace{-0.1cm}
\begin{split}
&\frac{\partial \mathcal L}{\partial y_i} =
 2\sum_{j\in\mathcal U_i}  \lambda_j a_j  ({y_i} - {y_j})=0.
\end{split}
\end{equation}
Solving  \eqref{optimal_problem3eq0} to \eqref{optimal_problem3eq2} yields
\vspace{-0.1cm}
\begin{equation}\label{optimal_problem3eq3}
\vspace{-0.1cm}
P_{i,T + 1}= {\left(\frac{2}{m+3}\sum_{j\in\mathcal U_i}  \lambda_j\right)^{\frac{m+3}{m+1}}},
\end{equation}
\begin{equation}\label{optimal_problem3eq3_2}
x_i=
\frac{ \sum_{j\in\mathcal U_i}  \lambda_j a_j {x_j}}{ \sum_{j\in\mathcal U_i}  \lambda_j a_j},
y_i=
\frac{ \sum_{j\in\mathcal U_i}  \lambda_j a_j {y_j}}{ \sum_{j\in\mathcal U_i}  \lambda_j a_j}.
\end{equation}

Given $x_i$, $y_i$ and $P_{i,T + 1}$, the value of $\lambda_j$ can be determined by the gradient method \cite{zhaohui}.
The updating procedure is:
\vspace{-0.1cm}
\begin{equation}\label{optimal_problem3eq5}
\lambda_j=
 \lambda_j+ \gamma \left((({x_j} - {x_i}) ^2  +  ({y_j} - {y_i}) ^2  + H^2)a_j -{P_{i,T + 1}^{\frac{2}{m+3}}}  \right),
\end{equation}
where $\gamma$ is a dynamic step-size. With regards to the optimality, we formulate the original optimization problem as a convex problem and, hence, it can always converge to the optimal solution.

The proposed algorithm used to solve problem in (\ref{optimal_problem}) is summarized in \textbf{Algorithm 2}. The complexity of the proposed algorithm lies in training a GRU-based prediction model and updating UAV location $(x_i,y_i)$.
A ground server can be deployed to train a GRU-based prediction model using \textbf{Algorithm 1}. Based on the trained model, the complexity of predicting the illumination distribution at the next time slot is $\mathcal O\left(1\right)$. In addition, the complexity of calculating $(x_i,y_i)$ of each UAV $i$ is $\mathcal O\left(L_i|\mathcal U_i|\right)$, where $L_i$ is the number of iterations of UAV $i$ until (\ref{optimal_problem3}) convergence and $|\mathcal U_i|$ is the number of users covered by UAV $i$. Therefore, the proposed algorithm can run independently on each UAV due to the linear algorithm complexity.

\begin{algorithm}[t]
\vspace{-0.05cm}
\footnotesize
\caption{The Overall Algorithm for Deploying UAVs.}
\begin{algorithmic}[1]
\STATE \textbf{Input:} A time series dataset of night remote sensing images. The set of locations of users in ${\cal{U}}$. Height $H$ of UAVs.
\STATE \textbf{Initialize:} The set of data rate requirement of users in ${\cal{U}}$. ${\cal{U}}_i$ served by each UAV $i$. Dual variables $\lambda_j$.
\FOR {$t = 1 \to T$}
\STATE EM algorithm for fitting GMM of illumination distribution at time $t$.
\STATE Obtain the optimal vector ${\bm{q}_{t}}$.
\STATE Add vector ${\bm{q}_{t}}$ to matrix $\boldsymbol{Q}$.
\ENDFOR
\STATE Input $\boldsymbol{Q}$ into \textbf{Algorithm 1} to predict the illumination distribution at time $T+1$, ${I_{T+1}(x,y)}$.
\FOR{{$i = 1 \to D$}}
\REPEAT
\STATE Update  transmission power $P_{i,T + 1}$ and UAV location $(x_i,y_i)$ according to (\ref{optimal_problem3eq3})-(\ref{optimal_problem3eq3_2}).
\STATE Update dual variables  $\lambda_j$, $j\in\mathcal U_i$ based on (\ref{optimal_problem3eq5}).
\UNTIL the objective function (\ref{optimal_problem3}) converges.
\STATE Calculate the transmit power $P_{i,T+1}$ based on the position of UAV $i$ being $(x_i,y_i,H)$ and the illumination distribution being ${I_{T+1}(x,y)}$.
\ENDFOR
\STATE \textbf{Output:} $P = \sum\limits_{i \in {\cal{D}}} {P_{i,T+1}} $.
\end{algorithmic}
\label{algorithm_2}
\vspace{-0.05cm}
\end{algorithm}

\section{Simulation Results and Analysis}
\label{sec:4}
For our simulations, a $20$~m $\times$ $20$~m square area is considered with $U=40$ uniformly distributed users and $D=4$ UAV. The downlink rate requirement $R_j$ of each user $j$ is generated randomly and uniformly over [0.5,1.5] Mbps. Other parameters are listed in Table \uppercase\expandafter{\romannumeral1}. The time series illumination data used to train GRU based prediction algorithm is a dataset of  average radiance composite nighttime remote sensing images, obtained from the Earth observations group (EOG) at NOAA/NCEI \cite{EOG}.
\begin{table}\footnotesize
\setlength{\belowcaptionskip}{0pt}
\setlength{\abovedisplayskip}{-15pt}
\newcommand{\tabincell}[2]{\begin{tabular}{@{}#1@{}}#1.6\end{tabular}}
\renewcommand\arraystretch{1}
\caption[table]{{System parameters}}
\centering
\begin{tabular}{|c|c|c|c|c|c|}
\hline
\!\textbf{Parameters}\! \!\!& \textbf{Value} &\! \textbf{Parameters} \!& \textbf{Value} \\
\hline
$\Phi $ & $120^\circ$ &  $\Psi_c$ & $120^\circ$ \\
\hline
$S$ & 1 cm$^2$  & $n_r$ & 1.5 \\
\hline
$H$ & 10~m &$\xi$& 1 \\
\hline
 $n _w$ & $1 \times {10^{ - 7}}$ & $\eta _r$ & $3 \times {10^{ - 7}}$\\
 \hline
 $D_h$ & $64$ & $D_q$ & 16  \\
 \hline
 $T$&358 & $K$ & 3\\
 \hline
 $\gamma$& 0.01 & $e$& 100 \\
\hline
\end{tabular}
\vspace{-0.5cm}
\end{table}

\vspace{-0.5cm}
\begin{figure}[htbp]
\centering
\subfigure[Actual illumination distribution.]{
\begin{minipage}[t]{0.4\linewidth}
\centering
\includegraphics[width=1.1in]{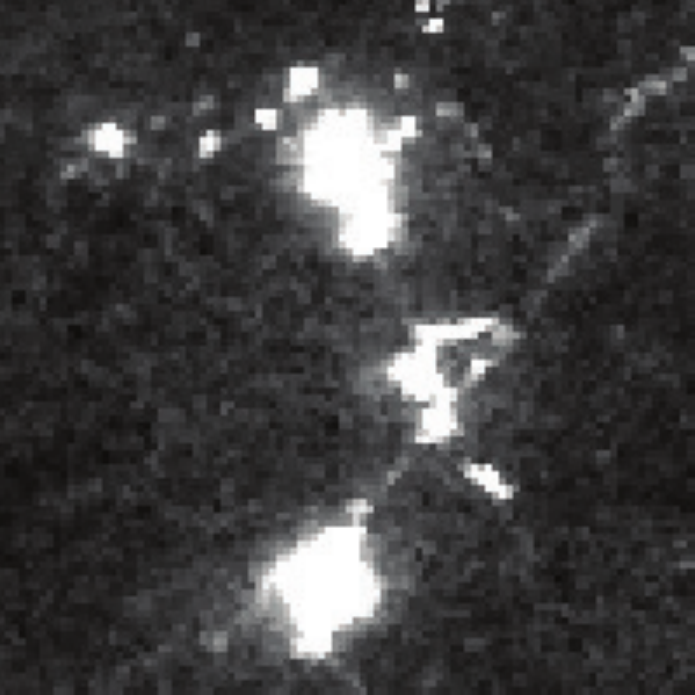}
\label{fig2a}
\end{minipage}%
}%
\subfigure[GMM modeled illumination distribution.]{
\begin{minipage}[t]{0.6\linewidth}
\centering
\includegraphics[width=1.9in]{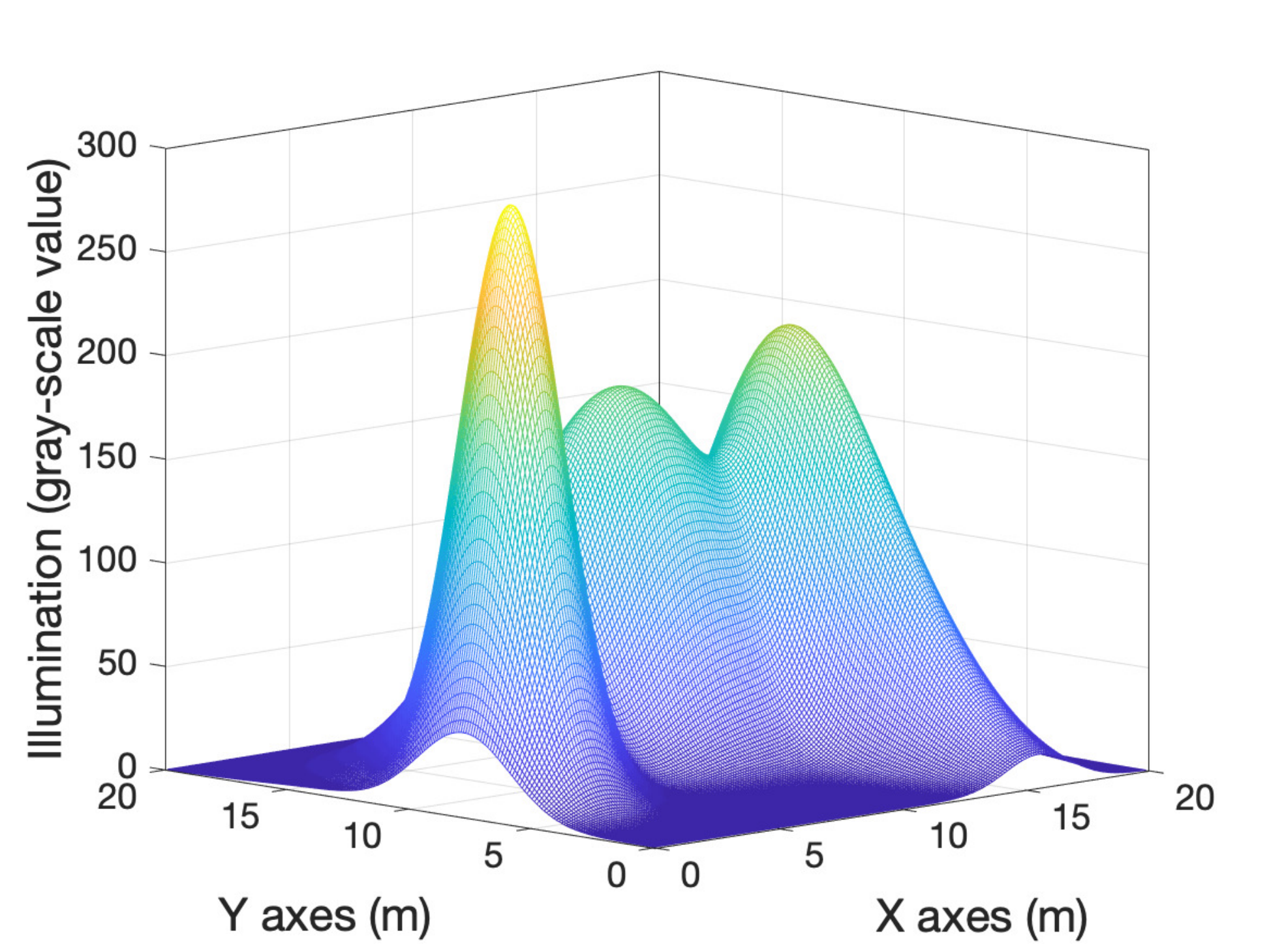}
\label{fig2b}
\end{minipage}%
}%
\centering
\caption{The illumination distribution of the target area.}
\label{fig2}
\vspace{-0.3cm}
\end{figure}

Fig. \ref{fig2} shows how the EM algorithm approximates the illumination distribution of the target area using a GMM model. Fig. \ref{fig2a} shows the actual illumination distribution and Fig. \ref{fig2b} shows the illumination distribution of the target area modeled by GMM. In Fig. \ref{fig2b}, we can see that EM algorithm uses 3 Gaussian components to model the actual illumination distribution. This is due to the fact that the area in Fig. 2(a) has only three lighting places.

\begin{figure}[t]
\centering
\setlength{\belowcaptionskip}{-0.45cm}
\vspace{-0.5cm}
\includegraphics[width=5.4cm]{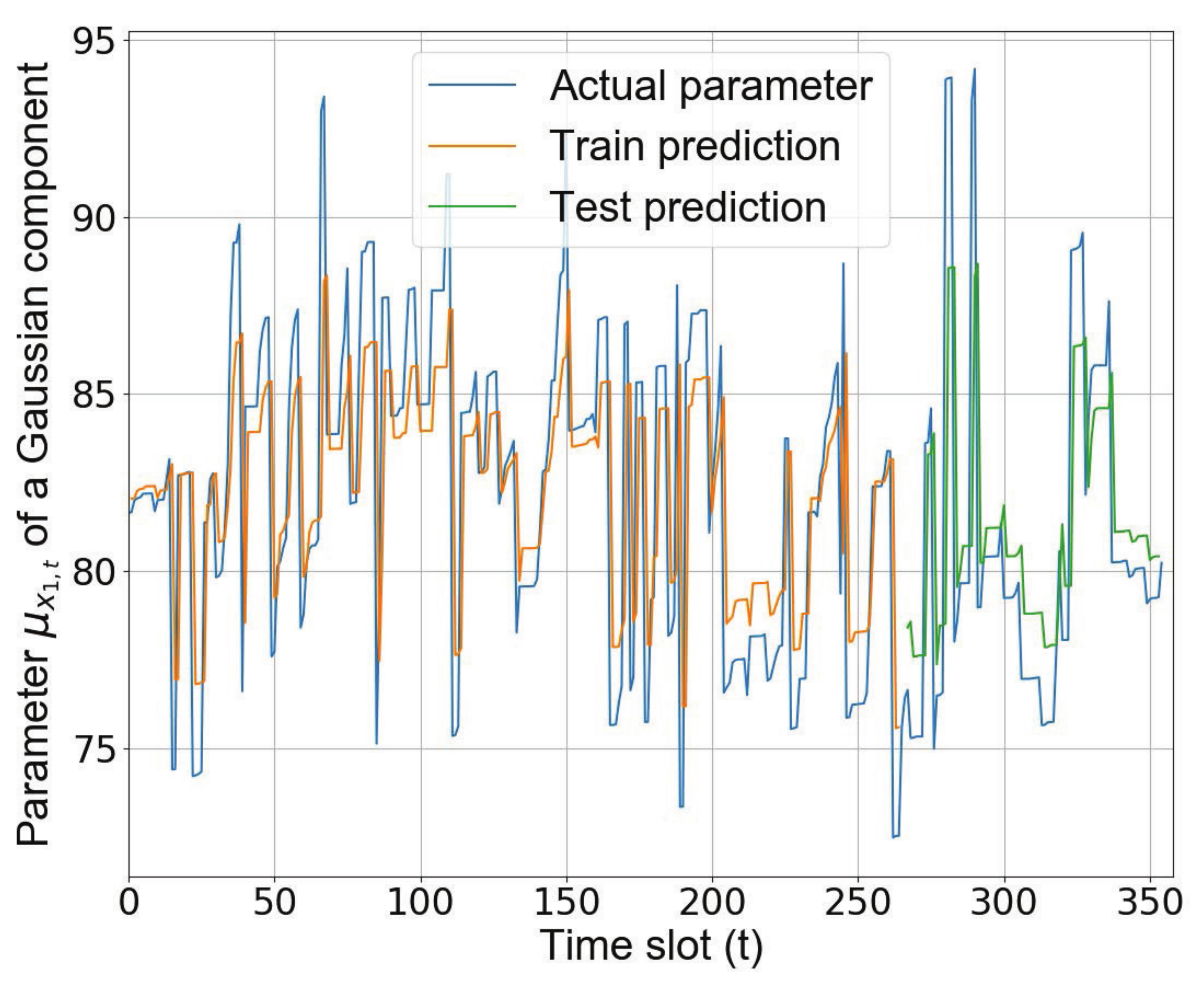}
\caption{Prediction accuracy of the illumination distribution.}
\vspace{0.2cm}
\label{fig3}
\end{figure}

Fig. \ref{fig3} evaluates the prediction accuracy of our GRU-based algorithm. 268 data samples are used to train the GRU-based model and the remaining 90 data samples are used to test the accuracy of the model. In Fig. \ref{fig3}, we can see that the proposed model can accurately predict the illumination distribution after 100 epochs. Over 5000 independent runs, the average root-mean-square error (RMSE) of training data prediction and test data prediction are 3.36 and 3.65, respectively. This is due to the fact that the GRU-based model can build a relationship between the prediction and the historical illumination distribution. Therefore, the GRU-based algorithm can predict the illumination distribution accurately.

\begin{figure}[t]
\centering
\setlength{\belowcaptionskip}{-0.45cm}
\vspace{-0.3cm}
\includegraphics[width=6.4cm]{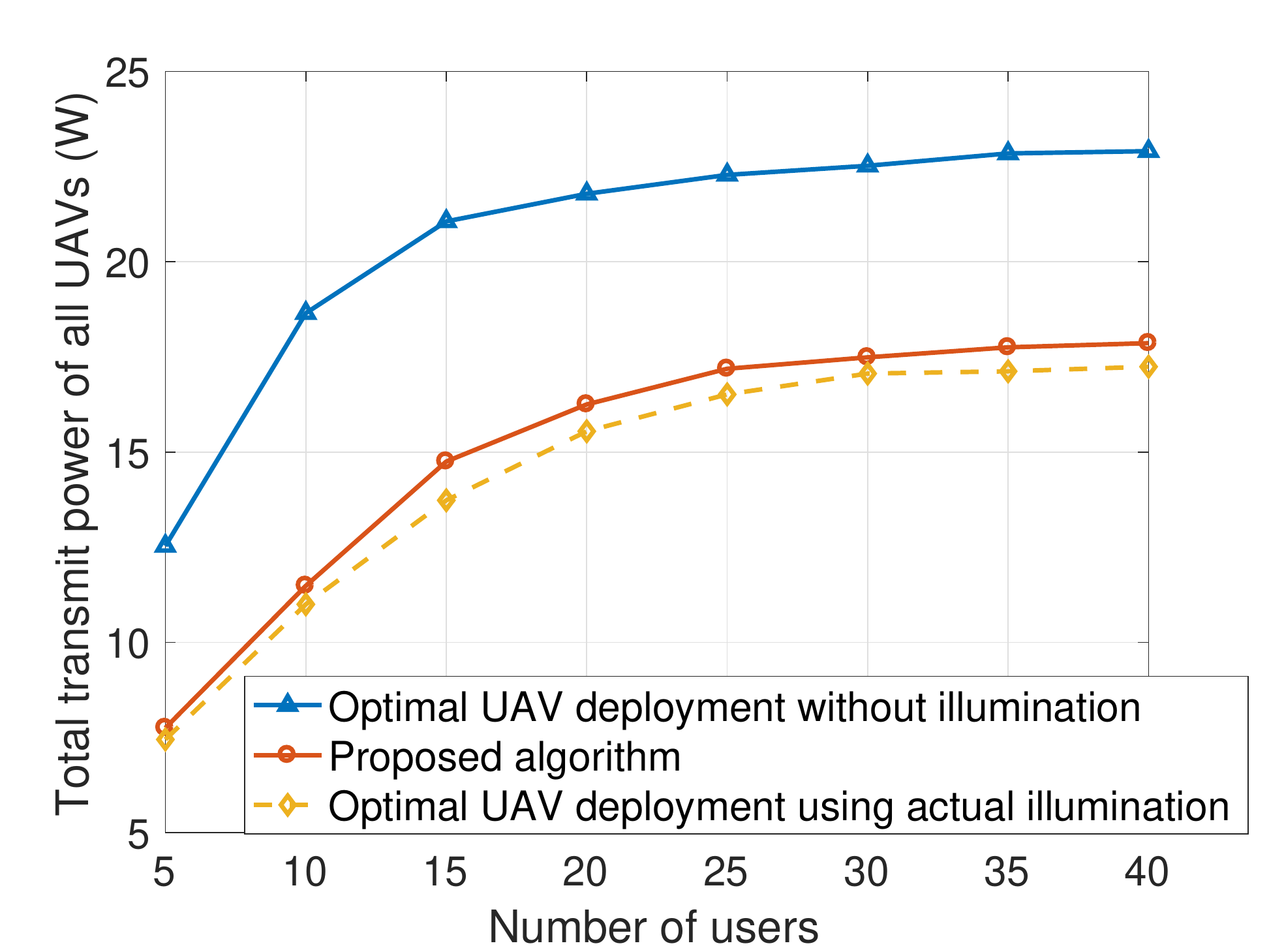}
\caption{The required sum power of UAVs as the number of users varies.}
\vspace{-0.1cm}
\label{fig4}
\end{figure}

Fig. \ref{fig4} shows how the transmit power used to meet the users' data rate and illumination requirements changes as the number of users varies. In this figure, we can see that the proposed algorithm achieves up to 22.1\% gain in terms of transmit power reduction compared to a conventional optimal UAV deployment without considering the illumination distribution. This is due to the fact that the power required by the users is related to the illumination of the service area. From Fig. \ref{fig4}, we can see that the proposed algorithm is closer to the UAV deployment optimization using actual illumination distribution and the gap between the two schemes is less than 3.4\%. This is because the proposed prediction algorithm can accurately predict the illumination distribution so as to optimize UAV deployment. Fig. \ref{fig4} also shows that, as the number of users increases, the performance gain of the proposed deployment becomes less significant. This is because when enough users are considered, the users will be uniformly distributed in the square and the optimal position of the UAV will be fixed.

\begin{figure}[t]
\centering
\setlength{\belowcaptionskip}{-0.45cm}
\vspace{-0.7cm}
\includegraphics[width=6.9cm]{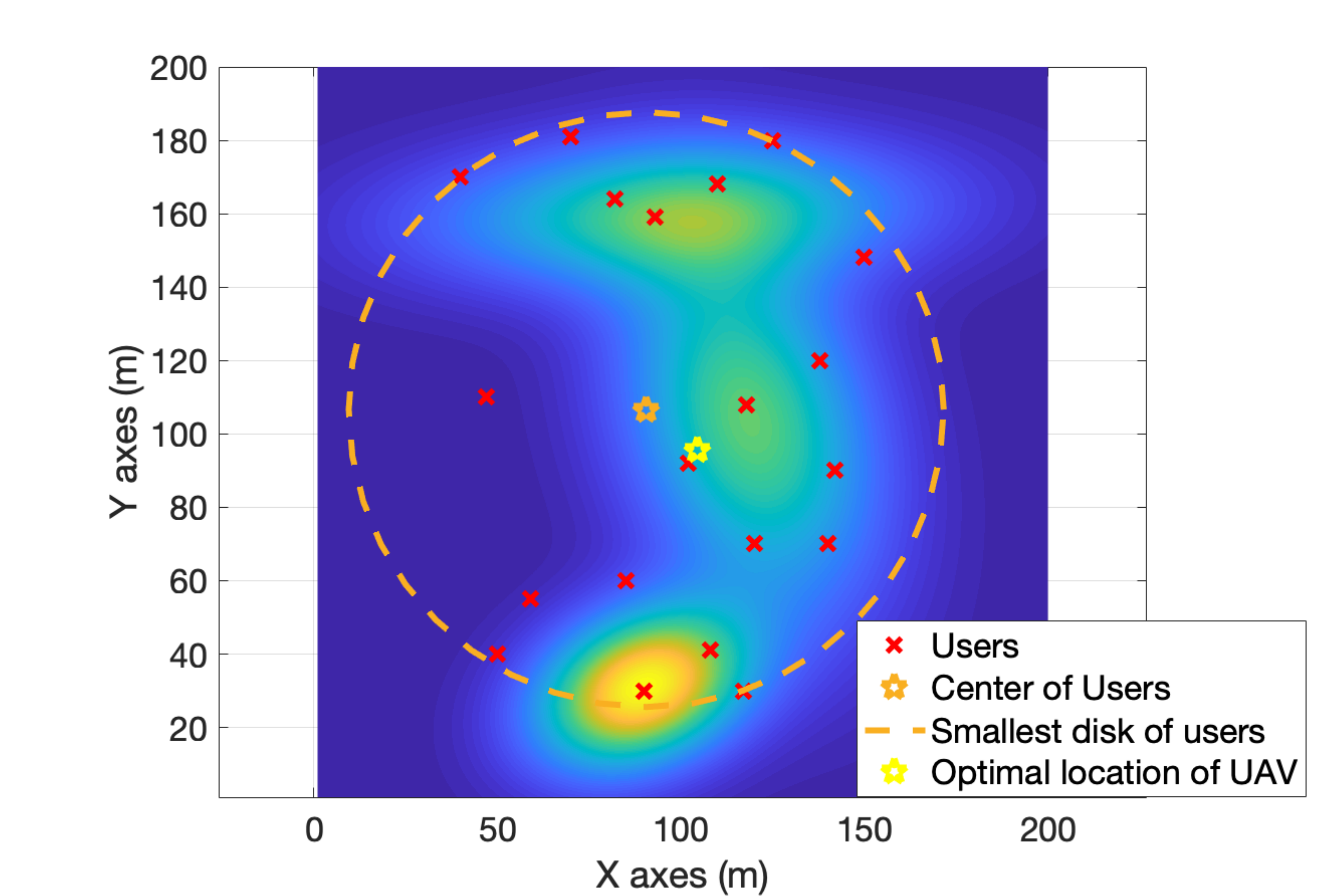}
\caption{Optimal deployment of UAVs using the proposed algorithm.}
\vspace{-0.2cm}
\label{fig5}
\end{figure}

In Fig. \ref{fig5}, we show an example of how the proposed algorithm can optimize the deployment of the UAV. From Fig. \ref{fig5}, we can see that the optimal location of the UAV without considering the illumination distribution is the center of all the users and the optimal location of the UAV obtained by the proposed algorithm are shifted to the area with strong illumination. This is due to the fact that the illumination increases the interference for VLC link and, hence, the users located in a bright area need more transmit power compared to those located in a dark area. Under the collective effect of all the users in the service area, the optimal UAV locations move towards the area with strong illumination to minimize the total transmit power.

\vspace{-0.05cm}
\section{Conclusion}
\label{sec:5}
In this paper, we have developed a novel UAV deployment framework for dynamically optimizing the locations of UAVs in a VLC-enabled UAV based network. We have formulated an optimization problem that seeks to minimize the transmit power while meeting the illumination and communication requirements of each user. To solve this problem, we have developed a GRU-based prediction algorithm, which can model the long-term historical illumination distribution and predict the future illumination distribution. We have then shown that the location optimization problem is convex and the optimal solution is obtained by using the dual method. Simulation results have shown that the proposed approach yields significant power reduction compared to conventional approaches.






%



\bibliographystyle{IEEEbib}
\def\baselinestretch{0.85}
\bibliography{illumination}

\end{document}